# Coherent control of collective atom phase for ultralong, inversion-free photon echoes

Byoung S. Ham
Center for Photon Information Processing, School of Electrical Engineering, Inha University
253 Yonghyun-dong, Nam-gu, Incheon 402-751, S. Korea
bham@inha.ac.kr

**Abstract:** To overcome fundamental limitations of the π optical pulse-induced population inversion and optical decay-caused short storage time in conventional photon echoes, a coherent control of collective atoms is studied for inversion-free, optical decay-halted photon echoes, where the constraint of photon storage time is now replaced by a spin population decay process. Using phase-controlled double rephasing, an inversion-free photon echo scheme is obtained, where no spontaneous or stimulated emission-driven quantum noise exists. Thus, the present method can be applied for ultralong quantum memories in quantum repeaters for long-distance quantum communications.
PACS numbers: 42.50.Md, 82.53.Kp

Over the last decade, quantum information processing has been widely investigated for potential applications of quantum computing and quantum communications. To overcome a limited transmission distance through an optical fiber, quantum communications beyond 100 km has been an ultimate goal. Using quantum repeaters [1], the limited transmission distance can be greatly extended, as with optical amplifiers in the present fiber-optic networks. In this case, the (quantum) information bandwidth is not degraded. Ultralong quantum memory is essential to work with quantum repeaters [1]. Here, the minimum storage time for global quantum communications is about one second [2]. However, such an ultralong storage quantum memory protocol does not exist.

Photon echoes have been intensively studied for all-optical information processing owing to their wide bandwidth, fast access time, and multimode capabilities in both time and space domains [3,4]. Photon echoes are based on reversibility of inhomogeneously broadened atom (or spin) phases, where the reversible atom phase is obtained by an optical π pulse-caused population swapping between two states. For quantum memory applications, however, photon echoes have been excluded as potential candidates due mainly to rephasing-caused population inversion, where spontaneous emission or a stimulated gain process deteriorates quantum fidelity. Obtaining optical decay time much shorter than a millisecond is another obstacle for quantum memories. Practically speaking, solid media have no atomic diffusion and pose a longer decay time compared with atomic vapors, where atoms must be tightly localized in space to avoid coherence loss for a certain period of time [5].

Recently several modified photon echo schemes have been demonstrated for quantum memories to resolve the population inversion problem in photon echoes [6-9]. However, the resultant storage time is still too short (less than a millisecond) to apply to global quantum communications. Here, an ultralong, noise-free photon echo scheme is proposed using coherent control of collective atom phases. For the ultralong, noise-free photon echoes, double rephasing is combined with optical locking originally proposed in resonant Raman echoes for ultralong photon storage [10]. In this case, both phase locked atoms as an inherent property of stimulated photon echoes [4] and controlling atom phases for the optical-spin coherence transfer play a key role [10,11].

Figure 1 shows an inversion-free, storage time-extended stimulated photon echo, where double rephasing and optical locking are combined for atom phase control. For Fig. 1, an inhomogeneously broadened lambda-type three-level optical system is considered, where D (*data*), W (*write*), R (*read*), and RR (*rephasing*) pulses are resonant for the transition of |1> (ground) to |3> (excited).

The optical locking pulses C1 and C2 are resonant to the transition of |2> − |3> for coherent population transfer [11]. Because spin decay rates are robust compared with optical counterparts, C1 functions as a storage time extension due to optical decay halt [12,13]. Unlike two-pulse photon echoes [3], the photon storage time extension by C1 in Fig. 1 is independent of optical phase decay as well as spin phase decay due to intrinsic atom phase locking in the stimulated photon echoes (discussed in Fig. 2).

The inset of Fig. 1(a) shows a corresponding pulse sequence, where D, W, and R are for a conventional stimulated photon echo. The pulse area of R ($\Phi_R=3\pi/2$), however, increases threefold over the conventional one ($\Phi_R=\pi/2$) to satisfy the phase recovery condition for optical locking (discussed in Figs. 3 and 4): $\Phi_R=(4n-1)\pi/2$; n=1,2,3... [10,11]. Unlike the optically locked echoes [10], the photon echo E1 in Fig. 1(a) is silent resulting in no photon echo generation due to absorptive (or negative) coherence (discussed in Fig. 4). Moreover, atom absorption by E1 is also prohibited under the inverted system. Therefore, the echo E1 does not alter system coherence.

The last pulse RR satisfies a two-pulse photon echo scheme with echo E1, where echo E2 is the rephased coherence of echo E1 with no population inversion. Thus,

Fig. 1(a) satisfies inversion-free, storage time extended photon echoes.

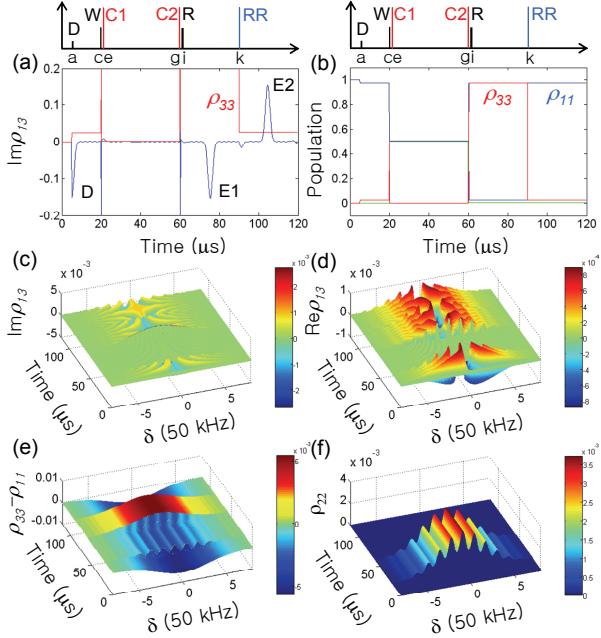

Fig. 1. Inversion-free storage time extended photon echo. (a) and (b). Optical coherence and population as a function of time, respectively. Pulse area of D, W, R, and RR is $0.1\pi$, $0.5\pi$, $1.5\pi$, and $\pi$, respectively. Inset: pulse sequence. $T_D(a)=5.0$; $T_W(c)=20.0$; $T_{C1}(e)=20.1$; $T_{C2}(g)=60.0$; $T_R(i)=60.1$; $T_{RR}(k)=90.0$ μs. Red: $\rho_{33}$. (c)-(f) Coherence evolution for all atoms as a function of detuning and time. The optical inhomogeneous width (FWHM) is 340 kHz. Each pulse duration is 100 ns except for D and R. D pulse area is $0.1\pi$. All decay parameters are set to zero.

Figures 1(c) - 1(f) show coherence evolutions of all excited atoms. In Figs. 1(c) and 1(e), the echo E2 at t=105 μs is without population inversion, while the (absorptive or negative) echo E1 at t=75 μs is with population inversion. The storage time extension is due to the population transfer by C1 to the spin state |2> [ ], where atom phase locking resulted from the W pulse is still effective for spin transitions (discussed in Fig. 2), even under spin inhomogeneous broadening as experimentally demonstrated [14]

For the calculations, nine time-dependent density matrix equations are numerically solved for an 800 kHz inhomogeneously broadened Gaussian distributed atom ensemble by dividing it into 161 subgroups. For visualization purposes, all decay rates are set at zero. Because the pulse duration of the *data* pulse D is as short as the inverse of the inhomogeneous broadening, the delay time of the *write* pulse W from D can also be kept short enough compared with optical phase decay time for nearly maximum coherence.

In Figs. 2 and 3, we discuss how the phase-locked atoms play an important role via optical-spin coherence transfer for storage time extension. We also discuss collective atom phase control in the optical locking processes using extra rephasing. The blue and red curves in Fig. 2(a) indicate D-pulse-excited coherence (Im$\rho_{13}$) and population ($\rho_{33}$) in a spectral domain at t=5.1 μs, respectively. Defining $\delta$ as a detuning of an atom group from the line center in the inhomogeneously broadened ensemble, each $\delta$–detuned atom evolves at a different speed ($\exp(\pm i\delta t)$; see Figs. 1(c) and 1(d)), resulting in a phase modulation (see the dotted curve for t=20.0 μs). The cyan curve is for the corresponding real part, Re$\rho_{13}$, which is $\pi/4$ phase shifted. The quantum optical information of the data pulse D is stored in this time-dependent optical phase modulation, i.e. atom phase grating. As rephasing-based photon echoes have been demonstrated for quantum memories [6-9], the stimulated photon echo should satisfy quantum memories, if the optical population decay-caused coherence loss is ignored [7,8].

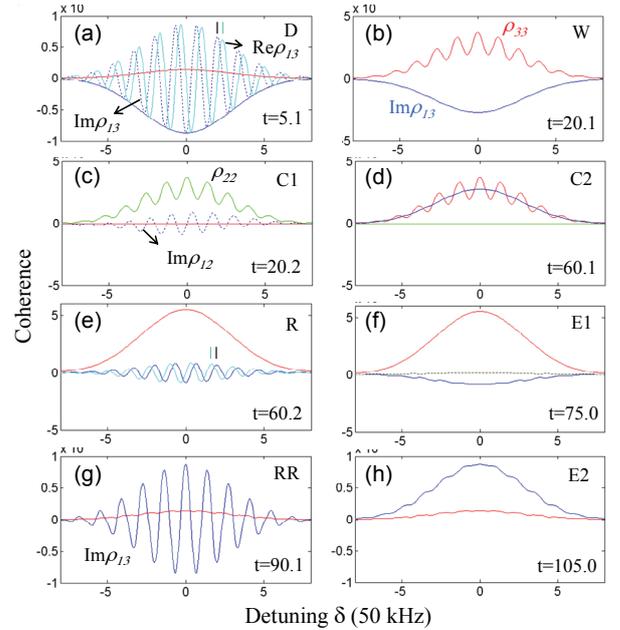

Fig. 2. Coherence versus detuning for Fig. 1. (a) t=5.1 μs for D excitation in the $0.1\pi$ pulse area. (b) t=20.1 μs for W excitation in the $\pi/2$ pulse area. (c) t=20.2 μs for C1 ($\pi$ pulse area). (d) t=60.1 μs for C2 ($\pi$ pulse area). (e) R rephasing with the $3\pi/2$ pulse area. (f) Silent echo E1 at t=75 μs. (g) RR rephasing with a $\pi$ pulse area. (h) Inversion-free echo E2 at t=105 μs. Blue: Im$\rho_{13}$; Red: $\rho_{33}$; Green: $\rho_{22}$; Cyan: Re$\rho_{13}$.

The stimulated photon echo scheme is obtained from the two-pulse photon echo scheme by simply dividing the $\pi$ optical rephasing pulse into two halves, W and R [4]. In Fig. 2(b), a half $\pi$ pulse (W) excites only half of each population in each state, resulting in a population grating for both $\rho_{11}$ (not shown) and $\rho_{33}$ (red curve). This population grating has the same modulation as the phase grating, where the phase grating becomes washed out by the W pulse (see also the bottom row of Fig. 3). At this stage, system coherence becomes locked from the phase decay process, resulting in atom phase locking. This phase locking is an ultimate property of the stimulated photon echoes. In such echoes, the population grating becomes blurry as optical population decay proceeds, resulting in decoherence.

A subsequently followed π pulse of C1, however, coherently transfers the population $\rho_{33}$ into the spin state |2> to halt the population decay-caused decoherence, where spin population grating $\rho_{22}$ (green) is an exact replica of the optical population grating $\rho_{33}$ as shown in Fig. 2(c). Via this complete population transfer, optical coherence is also transferred into spin coherence (see the dotted curve for Im$\rho_{12}$). At this stage, optical coherence $\rho_{13}$ is completely frozen unless spin population decay proceeds, producing optical locking. In Fig. 2(d), by the identical C2 pulse, the transferred atoms are returned to state |3>, recovering both optical population and coherence. However, a π phase shift results (see the inverted blue curve compared with that of Fig. 2(b)).

With the modified rephasing pulse R the population grating (red) is transferred back into the phase grating (blue), and rephasing of all excited atoms begins. Here, the 3π/2 *read* pulse R is used to compensate the optical locking pulse-induced π phase shift (see Fig. 4(d)) [10]. The rephased coherence as an echo E1 in Fig. 2(f), however, is absorptive (see the blue) with population inversion (red), resulting in a silent echo. This absorptive echo property was anticipated in Fig. 2(e), where no phase change occurs in Im$\rho_{13}$ (blue).

By the second rephasing pulse RR at t=90.0 μs, E1 is reversed for E2 (see Figs. 2(g) and 2(h)). Because RR functions as two-photon echoes, the system undergoes no population inversion. Thus, echo E2 in Fig. 2(h) is free from both spontaneous and stimulated emission processes. The wiggles in Figs. 2(f) and 2(h) are due to phase grating, where R-induced population swapping should be δ−dependent as a common feature of photon echoes..

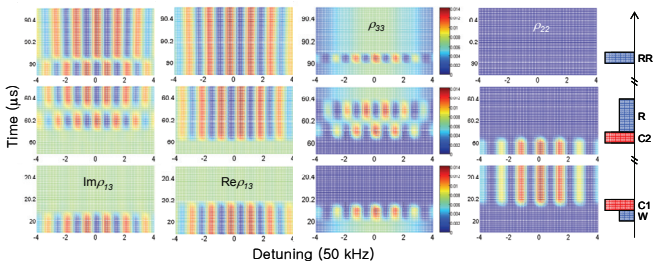

Fig. 3. Detailed coherence change by each pulse interaction as a function of time and detuning. All parameters are the same as those in Fig. 1, except the pulse area of D.

Figure 3 shows details of collective atom phase control. For maximum coherence excitation, a π/2 pulse area of D is used. The corresponding pulse sequence is the same as in Fig. 1 (see the far right column). As shown in the bottom row, the *write* pulse W functions as coherence transfer from phase grating ($\rho_{13}$) to population grating ($\rho_{33}$ as well as $\rho_{11}$ (not shown)). This population grating has nothing to do with the optical phase decay as discussed in Fig. 2(b). The optical locking pulse C1 results in coherence transfer from the optical state to the spin state via the population transfer as discussed in Fig. 2(c). Because the transferred atoms are phase locked, spin coherence can continue as long as $\rho_{22}$ decays out.

In the mid-row of Fig. 3, by C2, the spin population in state |2> is returned to state |3> with a total π phase shift, (see both Im$\rho_{13}$ and Re$\rho_{13}$). By the first π/2 pulse area of R at t=60.2 μs, however, the total accumulated phase shift becomes 2π, resulting in no rephasing and no echo: exp($i\delta t$) by D → exp($i\delta t$)$^{\exp(i\pi/2)}$=exp($-\delta t$) by W (phase evolution stops!)→ −exp($-\delta t$) by C1 and C2 (the imaginary part also experiences sign change)→ −exp($-\delta t$)$^{\exp(i\pi/2)}$=−exp($i\delta t$) by R(π/2). Thus, the additional π pulse area of R (at t=60.4 μs) is needed for rephasing: −exp($-i\delta t$). As discussed in Fig. 4, the following phase recovery condition must be satisfied for R:

$$\Phi_R = (4n-1)\pi/2, \qquad (1)$$

where $\Phi_R$ is the pulse area of *R*, and n is an integer.

However, this output leads to an absorptive photon echo (E1) under population inversion as discussed in Fig. 2(f). This condition explains why an additonal rephasing pulse RR is needed, where rephasing of E1 results in the inversion-free photon echo E2: exp($idt$)$^{\exp(i\pi)}$=exp($-i\delta t$); after E1, only Re$\rho_{13}$ has a sign change: −exp($-i\delta t$) → exp($i\delta t$). In other words, the final echo E2 is the exact coherence retrieval of E1 in the context of a conventional two-pulse photon echo. This fact leads to the important conclusion that the order of the inversion-free echo E2 is the same as that of data D (see Fig. 4(e)).

In Fig. 4, the atom phase control discussed in Figs. 1~3 is speculated using a Bloch vector model. Figure 4(a) shows the coherence evolution of a detuned atom group at $\delta$=10 kHz, where coherence evolution b-c is retrieved by RR.

Figure 4(b) represents a Bloch model for Fig. 4(a). The D-excited atomic coherence at point b is retrieved by R resulting in photon echo E1 at the bottom of the figure very near point b. The coherence of echo E, however, is negative leading to a silent echo under the population inversion (see Fig. 1(a)). By the additional rephasing pulse RR, echo E1 is retrieved for echo E2 (see the 'x' mark on the top curve), without population inversion as discussed in Figs. 1(a) and 2(h). As understood in Fig. 4(b), the RR must follow the first echo E1; otherwise, it does not satisfy the two-pulse photon echo scheme for echo E2.

Figures 4(c) and 4(d) represent the atom-phase control by R to support the phase recovery condition in Eq. (1). As shown in Fig. 4(d), for a complete retrieval of E1 the $\pi$ pulse area of R (blue region between points x and y) must be added to give a $\pi$ phase shift in total. As discussed in Fig. 3, where C1 and C2 result in a $\pi$ phase shift to $\rho_{13}$, the value of Im$\rho_{13}$ and Re$\rho_{13}$ at point x must be opposite to that at point c: exp($i\delta t$) at c → −exp($i\delta t$) at x. This condition represents no rephasing and no echo (see Fig. 4(f)).

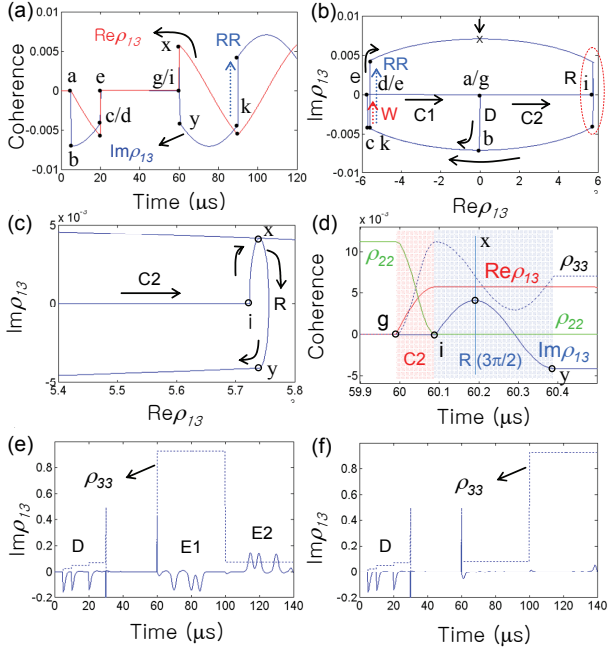

Fig. 4. (a) Coherence evolution for $\delta$=10 kHz. (b) Bloch vector description for (a). (c) and (d) Expanded features for the R interaction (see the red dotted circle in (b)). (e) Three consecutive data pulse for (a). (f) For $\pi/2$ pulse area of R.

In Fig. 4(e) the present inversion-free photon echo scheme is described for multiple data pulses. As shown, the order of final echo E2 is the same as that of the data pulses. This normally ordered echo sequence gives a practical benefit to multimode wide bandwidth photon information processing. This normally ordered echo sequence is the direct result of the double rephasing by adding RR.

Figure 4(f) supports the phase recovery condition with $\pi/2$ R, which violates Eq. (1). As a result, there is no photon echo due to no rephasing as discussed in Fig. 4(d). Echo E2 is also cancelled because of no E1.

In conclusion, coherent atom phase control in a collective atom ensemble was studied for ultralong, inversion-free photon echoes. Storage time extension is due to the inherent property of phase locking in stimulated photon echoes via optical-spin coherence transfer. Using optical phase locking, the transferred spin coherence becomes independent from spin dephasing, where the present scheme can be applied for ultralong quantum memories applicable to quantum repeaters in long-distance quantum communications.

This work was supported by the Creative Research Initiative Program (grant no. 2011-0000433) of the Korean Ministry of Education, Science and Technology via the National Research Foundation.